%% file: SMP-13-012_temp.tex
\pdfoutput=1

\documentclass[11pt,twoside,a4paper,cmspaper,final,collab]{cms-tdr}
\def\svnVersion{375613}\def\cmsCernNoTag{CERN-EP/2016-093}\def\cmsCernDate{\today}\def\cmsMessage{Published in the Journal of High Energy Physics as \href{http://dx.doi.org/10.1007/JHEP11(2016)147}{\doi{10.1007/JHEP11(2016)147.}}}

\begin{document}\cmsNoteHeader{SMP-13-012}

\hyphenation{had-ron-i-za-tion}
\hyphenation{cal-or-i-me-ter}
\hyphenation{de-vices}
\RCS$Revision: 371179 $
\RCS$HeadURL: svn+ssh://svn.cern.ch/reps/tdr2/papers/SMP-13-012/trunk/SMP-13-012.tex $
\RCS$Id: SMP-13-012.tex 371179 2016-10-19 14:42:12Z alverson $
\newlength\cmsFigWidth
\ifthenelse{\boolean{cms@external}}{\setlength\cmsFigWidth{0.85\columnwidth}}{\setlength\cmsFigWidth{0.4\textwidth}}
\ifthenelse{\boolean{cms@external}}{\providecommand{\cmsLeft}{top}}{\providecommand{\cmsLeft}{left}}
\ifthenelse{\boolean{cms@external}}{\providecommand{\cmsRight}{bottom}}{\providecommand{\cmsRight}{right}}
\newcommand{\mjj}{\ensuremath{m_{\mathrm{jj}}}\xspace}
\newcommand{\ptjone}{\ensuremath{{\pt}{}^{\mathrm{j}1}}\xspace}
\newcommand{\ptjtwo}{\ensuremath{{\pt}{}^{\mathrm{j}2}}\xspace}
\newcommand{\etaj}{\ensuremath{\eta^{\mathrm{j}}}\xspace}
\newcommand{\yjone}{\ensuremath{y_{\mathrm{j}1}}\xspace}
\newcommand{\yjtwo}{\ensuremath{y_{\mathrm{j}2}}\xspace}
\newcommand{\x}{\ensuremath{\phantom{0}}}

\cmsNoteHeader{SMP-13-012}
\title{Measurement of electroweak production of a \texorpdfstring{$\PW$}{W} boson and two forward jets in proton-proton collisions at \texorpdfstring{$\sqrt{s}=8\TeV$}{sqrt(s) = 8 TeV}}

\date{\today}

\abstract{
A measurement is presented of the cross section for the electroweak production of a $\PW$ boson in association with two jets in proton-proton collisions at a center-of-mass energy of 8\TeV. The data set was collected with the CMS detector and corresponds to an integrated luminosity of 19.3\fbinv. The measured fiducial cross section for \PW~bosons decaying to electrons or muons and for $\ptjone > $ 60\GeV, $\ptjtwo > $ 50\GeV, $\abs{\etaj} < $ 4.7, and $\mjj > $ 1000\GeV is $0.42 \pm 0.04\stat \pm 0.09\syst \pm 0.01\lum \unit{pb}$. This result is consistent with the standard model leading-order prediction of $0.50 \pm 0.02\,(\text{scale}) \pm 0.02\,(\mathrm{PDF})\unit{pb}$ obtained with {\MADGRAPH}5\_a\MCATNLO~2.1 interfaced to \PYTHIA 6.4. This is the first cross section measurement for this process.
}

\hypersetup{%
pdfauthor={CMS Collaboration},%
pdftitle={Measurement of electroweak production of a W boson and two forward jets in proton-proton collisions at sqrt(s)=8 TeV},%
pdfsubject={CMS},%
pdfkeywords={CMS, physics, electroweak, W boson}}

\maketitle

\section{Introduction}

The production of a $\PW$~or $\PZ$ boson in association with two jets via the $t$-channel exchange of an electroweak gauge boson (EW production) plays an important role in testing the gauge sector of the standard model (SM),
in particular, aspects of gauge boson self interactions.
This process is characterized by the presence of two jets with a large separation in rapidity~\cite{Rainwater:1996ud,Khoze:2002fa}.
Experimental studies of this topology are challenging because of large backgrounds
and require a precise understanding of extra quark and gluon emissions computed in quantum chromodynamics (QCD)~\cite{Chatrchyan:2013jya}.

\begin{figure*}[tbh]
    \includegraphics[width=0.33\textwidth]{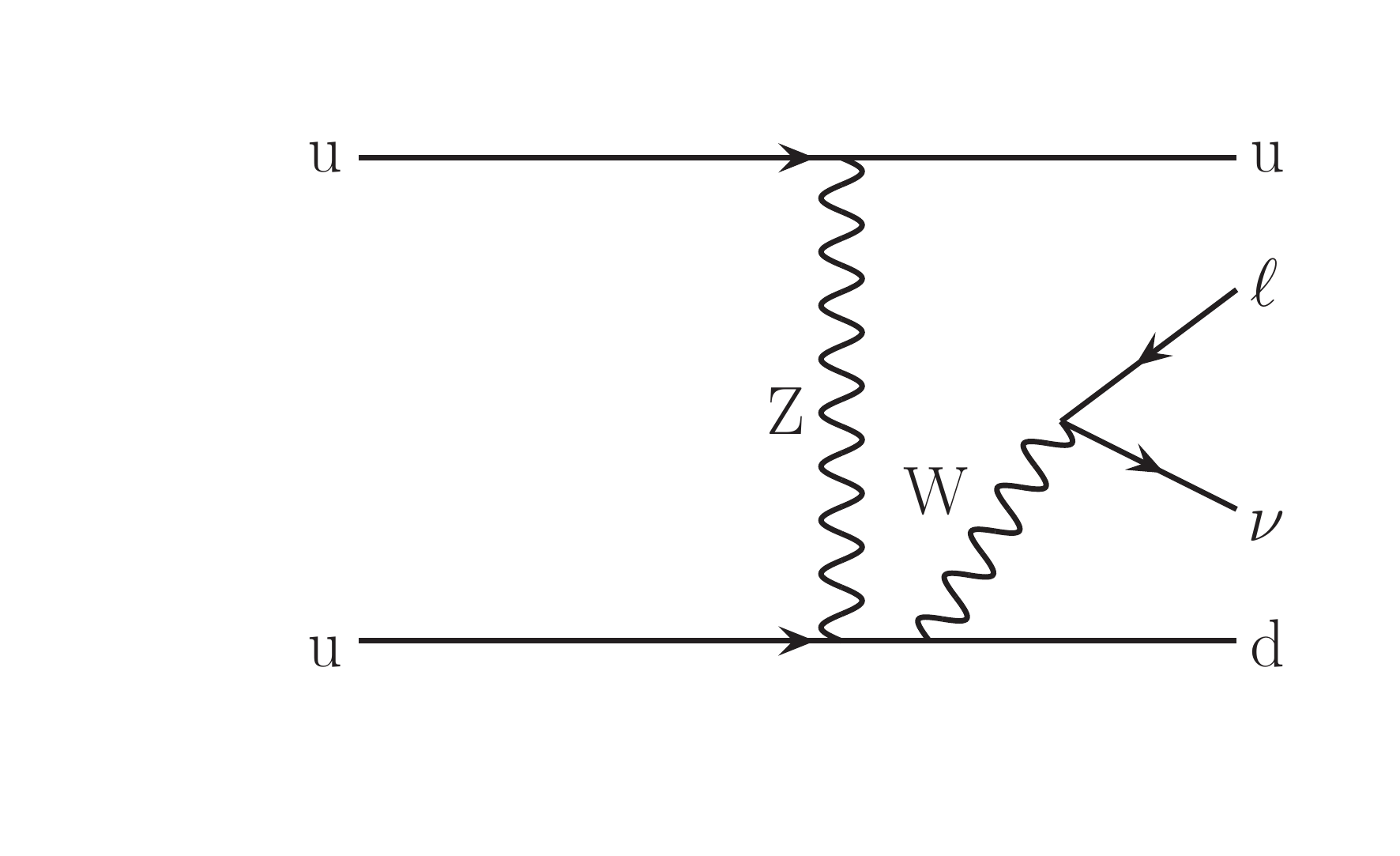}
    \includegraphics[width=0.33\textwidth]{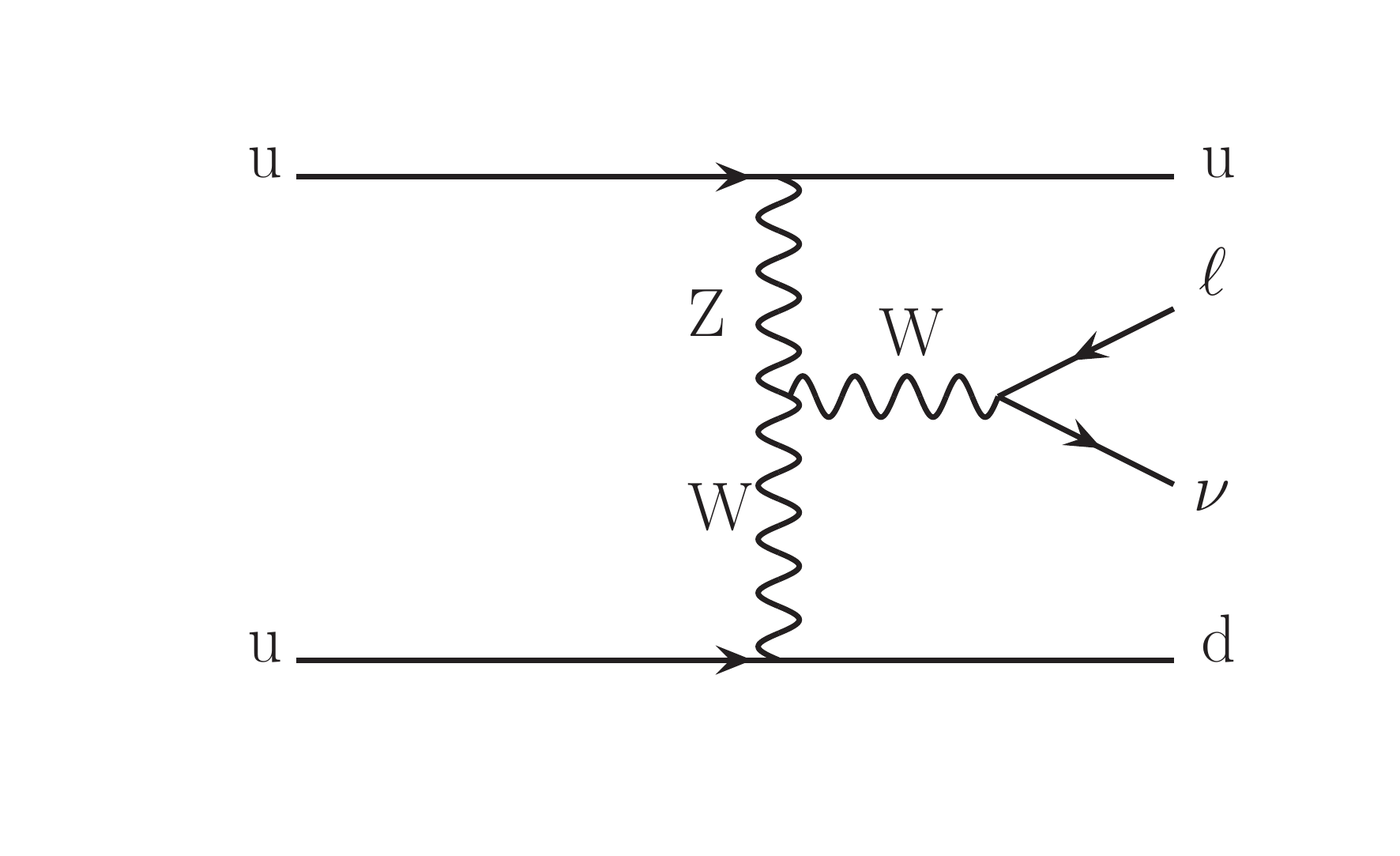}
    \includegraphics[width=0.33\textwidth]{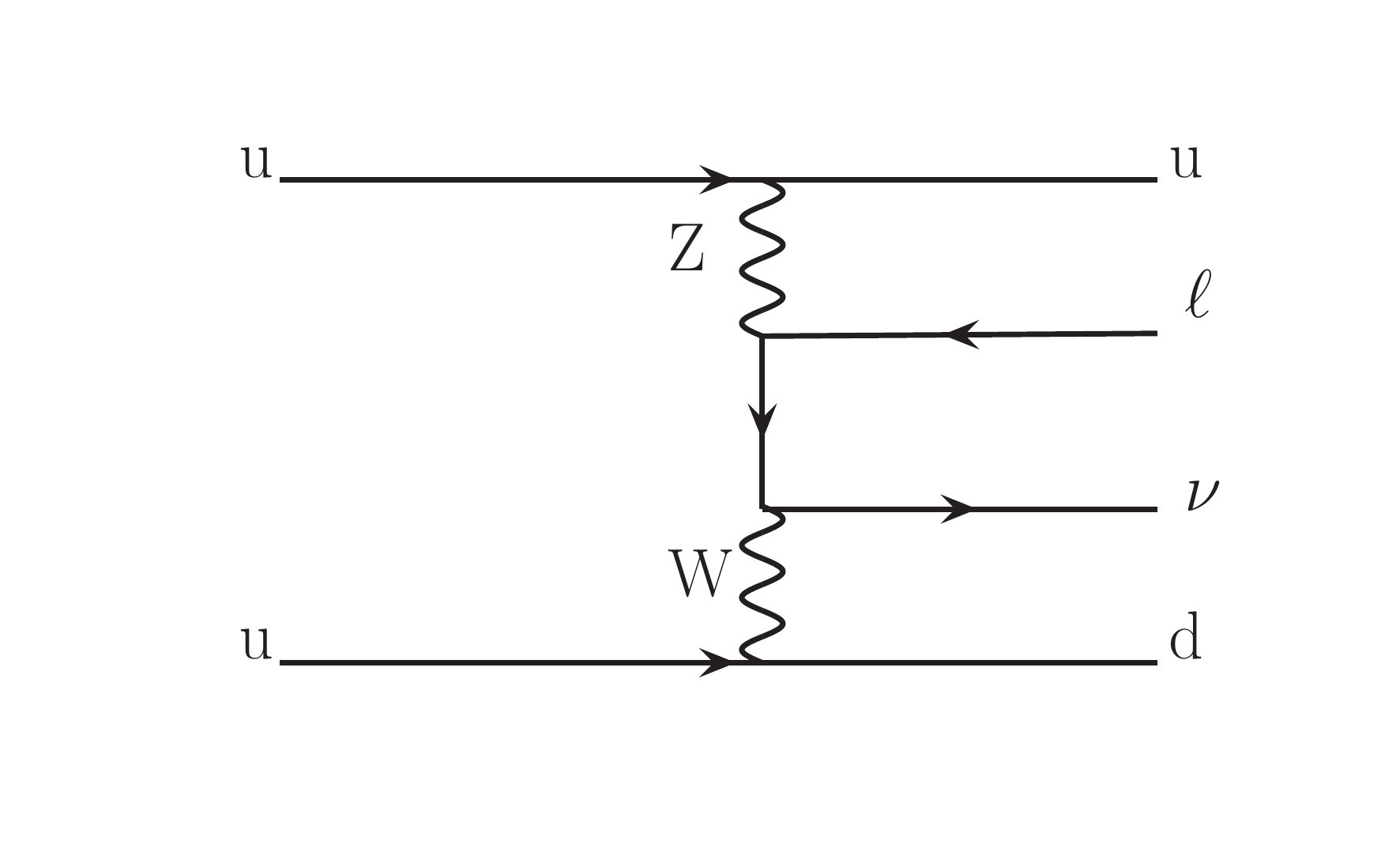}
	 \caption{\label{fig:wz_feynman} Representative diagrams for EW $\ell\nu \mathrm{jj}$ productions at the LHC: (left) bremsstrahlung, (center) VBF, and (right) multiperipheral processes.}
\end{figure*}

Three classes of EW diagrams for
$\ell\nu$ production
in association with two jets are shown in
Fig.~\ref{fig:wz_feynman}:
a $\PW$~boson radiating from a quark line (left), a $\PW$~boson produced through vector boson fusion (VBF) processes involving a $\PW$~boson and a $\PZ$~boson (center), and a multiperipheral diagram with no $s$-channel $\PW$~boson (right).
These diagrams represent the EW signal in this analysis.

The study of the EW $\PW$+2-jets process is part of a more general investigation of the SM VBF process.
These EW processes have been used to investigate the rapidity gaps at hadron colliders~\cite{Rainwater:1996ud,Khoze:2002fa}, as a probe of triple-gauge-boson couplings~\cite{Baur:1993fv,Eboli:2004gc}, and as a background to Higgs boson measurements in the VBF channel ~\cite{Rainwater:1998kj,Plehn:1999xi,Rainwater:1999sd,Kauer:2000hi}.

At the LHC, the EW production of a $\PZ$ boson in association with forward and backward jets has been investigated by the CMS Collaboration
at a center-of-mass energy
of 7\TeV~\cite{Chatrchyan:2013jya} (8\TeV~\cite{Khachatryan:2014dea}).
The ATLAS Collaboration has published similar results
at $\sqrt{s} = 8$\TeV~\cite{Aad:2014dta}.
The EW production of events with a same-sign $\PW$~boson pair plus two jets
was recently studied by the ATLAS Collaboration~\cite{Aad:2014zda} and the CMS Collaboration~\cite{Khachatryan:2014sta} at 8\TeV.

This paper presents a measurement of the EW $\PW$+2-jets production cross section.
The fiducial cross section is calculated
for $\PW$~bosons decaying to electrons or muons and for $\ptjone > 60$\GeV, $\ptjtwo >50$\GeV, $\abs{\etaj} <  4.7$, and $\mjj >  1000$\GeV.
The interference term between EW $\PW$+2-jets and QCD $\PW$+2-jets is neglected in the calculation of the fiducial cross section, and is considered as a source of systematic uncertainty.
The data set corresponds to an integrated luminosity of 19.2 (19.3)\fbinv collected by the CMS experiment in the electron (muon) channel at $\sqrt{s}=8\TeV$.

\section{The CMS detector}

The central feature of the CMS apparatus is a superconducting solenoid of 6\unit{m} internal diameter.
Within the solenoid volume are a silicon pixel and strip tracker, a lead tungstate crystal electromagnetic calorimeter (ECAL), and a brass and scintillator hadron calorimeter (HCAL), each composed of a barrel and two endcap sections.
Forward calorimeters extend the pseudorapidity~\cite{CMS:2010} coverage provided by the barrel and endcap detectors.
Muons are measured in gas-ionization detectors embedded in the steel flux-return yoke outside the solenoid.

The silicon tracker measures charged particles within the pseudorapidity range $\abs{\eta}< 2.5$. It consists of 1440 silicon pixel and 15\,148 silicon strip detector modules and is located in the 3.8\unit{T} field of the superconducting solenoid. For nonisolated particles of $1 < \pt < 10\GeV$ and $\abs{\eta} < 1.4$, the track resolutions are typically 1.5\% in \pt and 25--90 (45--150)\mum in the transverse (longitudinal) impact parameter \cite{TRK-11-001}

The electron momentum is estimated by combining the energy measurement in the ECAL with the momentum measurement in the tracker.
The momentum resolution for electrons with transverse momentum $\pt \approx 45 \GeV$ from $\Z \rightarrow \Pe \Pe$ decays ranges from 1.7\% for nonshowering electrons in the barrel region to 4.5\% for showering electrons in the endcaps~\cite{CMS:EGM-13-001}.
The electron objects in the transition region between the barrel and endcap
($1.44<\abs{\eta}<1.57$) are less precise.

Muons are measured in the range $\abs{\eta}< 2.4$, with detection planes made using three technologies: drift tubes, cathode strip chambers, and resistive plate chambers. Matching muons to tracks measured in the silicon tracker results in a relative $\pt$ resolution for muons with $20 <\pt < 100\GeV$ of 1.3--2.0\% in the barrel and better than 6\% in the endcaps. The \pt resolution in the barrel is better than 10\% for muons with \pt up to 1\TeV~\cite{Chatrchyan:2012xi}.

The HCAL, when combined with the ECAL,
measures jets with energy resolution amounting typically to 15\% at 10\GeV, 8\% at 100\GeV, and 4\% at 1\TeV.
The forward calorimeter modules extend the coverage of hadronic jets to $\abs{\eta} = 5.0$.

The missing transverse momentum vector \ptvecmiss is defined as the projection on the plane perpendicular to the beams of the negative vector sum of the momenta of all reconstructed particles in an event. Its magnitude is referred to as \ETmiss.
The CMS detector is nearly
hermetic, allowing for precise measurements of $\MET$.

The first level of the CMS trigger system, composed of custom hardware processors, uses information from the calorimeters and muon detectors to select the events of interest in a fixed time interval of less than 4\mus. The high-level trigger processor farm further decreases the event rate from around 100\unit{kHz} to
less than 1\unit{kHz}, before data storage.

A more detailed description of the CMS detector, together with a definition of the coordinate system used and the relevant kinematic variables, can be found in Ref.~\cite{CMS:2010}.

\section{Simulated samples}
\label{sec:SimulatedSamples}

Signal and background samples are simulated using standard packages.
The Monte Carlo (MC) event generator {\MADGRAPH}5\_a\MCATNLO~2.1~\cite{aMCatNLO}
is used to simulate the EW \PW$(\to \Pe\nu,\, \Pgm\nu,\, \tau\nu)$+2-jets
events.
Alternative
EW $\PW (\to \ell\nu)$
samples for systematic studies, where $\ell = \Pe$, $\Pgm$, as well as QCD induced $\PW$+jets, $\ttbar$, and Drell--Yan (DY) background events are simulated using
{\MADGRAPH}5~\cite{madgraph5}.
Single top quark production is modeled with \POWHEG
1.0~\cite{Nason:2004rx,POWHEG,Alioli:2010xd,Alioli:2009je,Re:2010bp}.
Diboson samples ($\PW\PW$, $\PW\PZ$, $\PZ\PZ$) are generated with
\PYTHIA~6.4~\cite{Sjostrand:2006za}.
All samples are generated using the
CTEQ6L1~\cite{Pumplin:2002vw}
parton
distribution function (PDF) set,
except for the \POWHEG single top quark sample,
for which the CTEQ6M~\cite{Pumplin:2002vw}
PDF set is used.
The parton showering and matching, hadronization, and underlying event
simulation for all samples are performed by \PYTHIA~6,
with the parameters of the underlying
event set to the Z2* tune~\cite{PythiaTuneZ2, Chatrchyan:2013gfi}.
The \TAUOLA 2.7 generator~\cite{Jadach1991275} is used to simulate $\tau$ lepton decays.
For systematic uncertainty studies the
alternative
signal sample is also
interfaced with {\HERWIGpp}~\cite{HERWIG},
which has different parton shower and hadronization models than \PYTHIA.

The QCD induced $\PW$+jets events, which is the main background, are generated with up to four partons using matrix element (ME) calculations.
The ME--parton-shower
matching scale
is taken to be 10\GeV~\cite{Hoche:2006ph}, and
the factorization and renormalization scales are both dynamically set to
$( M_{\PW}^2 + p_{\mathrm{T},\PW}^2 + \Sigma p_{\mathrm{T,j}}^2 )^{1/2}$,
where $M_{\PW}=80.4\GeV$ is the $\PW$~boson mass~\cite{Agashe:2014kda} and $\Sigma p_{\mathrm{T,j}}^2$ is the sum over the generated jets. The signal events are generated
with the same factorization and renormalization scales.

Alternative \ttbar\ samples are generated with \POWHEG~1.0
and \MCATNLO~3.4~\cite{MCatNLO1,MCatNLO2}. The \ttbar\ sample generated with \POWHEG~1.0
is interfaced with \PYTHIA 6.
The PDF set used for this sample is CT10~\cite{Lai:2010vv}.
The \ttbar\ sample generated with \MCATNLO~3.4 is interfaced with \HERWIG
~\cite{Corcella:2000bw,Corcella:2002jc}.
The PDF set used for this sample is CTEQ6M.

The cross sections for the signal samples are calculated at leading order (LO) using {\MADGRAPH}5\_a\MCATNLO~2.1.
The cross sections for the QCD $\PW$+jets and DY samples are normalized to the next-to-next-to-leading-order (NNLO) prediction calculated with \FEWZ 3.1~\cite{FEWZ}.
A data-driven method is used to normalize the QCD $\PW$+jets sample as described in Section~\ref{sec:Background_signal_estimation}.
The cross section for inclusive \ttbar\ sample is normalized to the next-to-next-to-leading-logarithm prediction from the \textsc{Top++}v2.0 generator~\cite{Czakon:2011xx}.
The cross section for the single top quark
production process is obtained from \textsc{Hathor} v2.1~\cite{Aliev:2010zk, Kant:2014oha}, which is accurate up to next-to-leading-order (NLO).
The cross sections for diboson samples are normalized to the NLO prediction calculated with MCFM 6.6~\cite{MCFM}.

A \GEANTfour-based simulation~\cite{GEANT4} of the CMS
detector is used in the production of all simulated
samples.
Additional proton-proton interactions within a bunch crossing (pileup)
are added to the simulation to match those observed in data.
During this data-taking period the mean number of
interactions per bunch crossing was 21.
Simulated events are corrected for efficiency differences relative to data using a tag-and-probe method~\cite{VBTF}.

\section{Event reconstruction and selection}
\label{sec:EventSelection}

A common event reconstruction algorithm and selection criteria are applied
to data and simulated events.
The event signature is an isolated lepton (electron or muon), two jets, and \MET.
Candidate events are collected with single-lepton triggers, which require an isolated electron (muon) with a \pt threshold of 27 (24)\GeV.
The overall trigger efficiency is 90\% (94\%) for the electron (muon) data,
with a small dependence on \pt and $\eta$.

The analysis relies on a particle-flow (PF) technique~\cite{CMS-PAS-PFT-09-001,CMS-PAS-PFT-10-001} that reconstructs various particles in the event (charged and neutral hadrons, electrons, muons, and photons) by optimally combining information from various CMS subdetectors.

Electrons are reconstructed from the combination of the tracker and the corresponding ECAL cluster information,
and must pass electron identification requirements according to a multivariate identification technique~\cite{CMS:EGM-13-001}.
Muons are reconstructed by fitting trajectories based on hits in the silicon tracker and in the outer muon system.
Lepton candidates are required to originate from the primary
vertex of the event, which is chosen to be the vertex with the highest value of
$\sum \pt^2$, where the sum is performed over all associated charged particle tracks.

Charged leptons from
$\PW$ boson decays are expected to
be isolated from other activity in the event.
Leptons
are required to fulfill a
requirement on their relative isolation,
which is defined as the ratio of the $\pt$ sum of all other PF candidates reconstructed in a cone $\Delta R = \sqrt{\smash[b]{(\Delta\eta)^2+(\Delta\phi)^2}} = 0.3$ $(0.4)$ around the candidate electron (muon) to the $\pt$ of the candidate, and is corrected for contributions from pileup.
The $\Delta\eta$ and $\Delta\phi$ are the differences in pseudorapidity and in azimuthal angle, respectively.
For electrons, the isolation selection is tuned together with the multivariate identification requirement to give an signal efficiency of 80\% independent of $\eta$.
Electrons are required to have a relative isolation smaller than 0.11, 0.18, and 0.15 for $\abs{\eta}$ ranges of 0.00--0.80, 0.80--1.48, and 1.48--2.50.
Muons are required to have a relative isolation smaller than 0.12 over the entire $\abs{\eta} < 2.1$ range used in the analysis.
The efficiency of the selection requirements for muons originating from $\PW$ boson decays is approximately 96\%.
To reduce the contribution from the DY production, the events are required to have no more than one isolated lepton in the final state.

Jets are reconstructed from PF candidates using the
anti-$\kt$
clustering algorithm~\cite{antikt,fastjetmanual} with a distance parameter 0.5.
Charged particles not originating from the primary vertex are not
considered for jet clustering~\cite{Cacciari:2008gn}. Jets from pileup are identified and removed with a pileup jet identification algorithm~\cite{CMS-PAS-JME-13-005}, based on both vertex information and jet shape information. A jet quality requirement, primarily based on the energy ratio between the charged and neutral hadrons within the jet cone~\cite{QCD-11-004JetIso}, is also applied in order to remove jets originating from calorimeter noise.
Jets overlapping
within $\Delta R =0.3$
with identified leptons
are not considered.

Pileup collisions and the underlying event can contribute to the energy of the reconstructed jets.
A correction based on the projected area of a jet on the front face of the calorimeter is used to subtract the extra energy deposited in the jet coming from
pileup~\cite{fastjet1,Cacciari:2008gn}. Furthermore, jet energy corrections are
applied to account for the nonlinear energy response of the calorimeters
and for other instrumental effects.
These corrections are based on in situ measurements using dijet,
$\gamma$+jet, and $\PZ$+jet data samples~\cite{Chatrchyan:2011ds}.

Electrons (muons) are required to have \pt
greater than 30 (25)\GeV and $\abs{\eta}<$ 2.5 (2.1).
The electrons found in the transition region between the barrel and endcap ($1.44<\abs{\eta}<1.57$) are not considered.
Events are required to have at least two jets with $\pt > 60$\GeV (leading) and $\pt > 50$\GeV (subleading),
both with $\abs{\eta}<4.7$.

To measure the W boson momentum an accurate $\MET$ measurement is essential.
We use $\MET$ measured in the event
with the full PF reconstruction~\cite{Khachatryan:2014gga}
and require $\MET >$ 30 (25)\GeV in the electron (muon) channel
to distinguish the $\PW$ boson signal from multijet backgrounds.
The leptonically decaying $\PW$ boson
is reconstructed by combining the kinematic information from the selected lepton with
\ptvecmiss.
The unmeasurable $p_z$ component of the neutrino can be estimated by assuming that the lepton and the $\MET$ arise from a $\PW$ boson with the nominal mass of 80.4\GeV.
A quadratic equation for the neutrino $p_{z}$ component is obtained that can be solved up to a two-fold ambiguity.
In the case of two real solutions, the neutrino $p_{z}$ solution that is closer to the charged lepton $p_{z}$ is selected.
In the case of two imaginary solutions, the common real part of the solutions is selected.

To reduce the background from events that do not contain
$\PW \to\ell\nu$ decays, we require that the transverse
mass of the $\PW$ boson candidate exceed 30\GeV. The transverse
mass of the leptonically decaying W boson is defined as $\sqrt{\smash[b]{2\pt^{\ell} \MET [1-\cos(\Delta \phi_{\ell, \ptvecmiss})]}}$, where $\Delta \phi_{\ell, \ptvecmiss}$ is the azimuthal angle between the lepton and the $\ptvecmiss$ directions.
In order to further improve
the signal over background ratio, two additional requirements are used:
the Zeppenfeld variable~\cite{Rainwater:1996ud}, defined as $|y_{\PW} -(\yjone + \yjtwo)/2|$, where $y$ represents rapidity, must be less than 1.2;
the invariant mass $\mjj$ of the jet pair is required to be greater than 1000\GeV.

Table~\ref{tab:SELECTION} provides a summary of the selection requirements.

\begin{table}
	\begin{center}
		\caption{Summary of selection criteria.}
		\label{tab:SELECTION}
		\begin{tabular}{l l}
			\hline
			         $\PW \to \ell \nu$ Lepton requirements  & Jet requirements\\ \hline
			          Single lepton trigger & $\ptjone >60\GeV$, $\ptjtwo >50\GeV$\\
			          High-quality lepton ID and isolation  & $ |y_{\PW} -(\yjone + \yjtwo)/2| < 1.2$\\
			          Electron (muon) $\pt> 30$ (25)\GeV   &  $\mjj > 1000\GeV$\\
			          $\MET> 30$ (25)\GeV for electron (muon) channels & \\
			          $\PW$ transverse mass $> 30$\GeV & \\
			          Veto second lepton & \\
			\hline
		\end{tabular}
	\end{center}
\end{table}

\section{Background estimation and signal extraction}
\label{sec:Background_signal_estimation}

The EW $\PW(\to\tau\nu)$+2-jets events, with the $\tau$ leptons decaying to electrons or muons, have signatures similar to those for signal events and
they represent
4\% (5\%) of the signal sample in the electron (muon) channel as estimated from simulation.
When calculating the fiducial cross section we require
that the $\PW$~boson decays to $\Pe$~or $\Pgm$~only, so $\tau$ events are not included as part of the signal.

A boosted decision tree (BDT) technique is used to distinguish between signal and background events and an unbinned maximum-likelihood fit to the $\mjj$ distribution is used to extract the number of signal events as described below.

The BDT technique is implemented in the Toolkit for multivariate data analysis
(TMVA)~\cite{2007physics...3039H}.
The adaptive boost algorithm (AdaBoost) \cite{Freund:1997:DGO:261540.261549} is used in the BDT training, which gives larger weights for decision trees with lower misclassification errors.
For good BDT performance,
variables well-modeled by simulation are required. The following input variables are used:
lepton $\eta$, $\Delta \eta$ between jets, $\Delta \eta$ and $\Delta \phi$ between the $\PW$~boson and each jet, and the $\PW$~boson \pt.

\begin{figure}[tb]
\centerline{
\includegraphics[width=.49\textwidth]{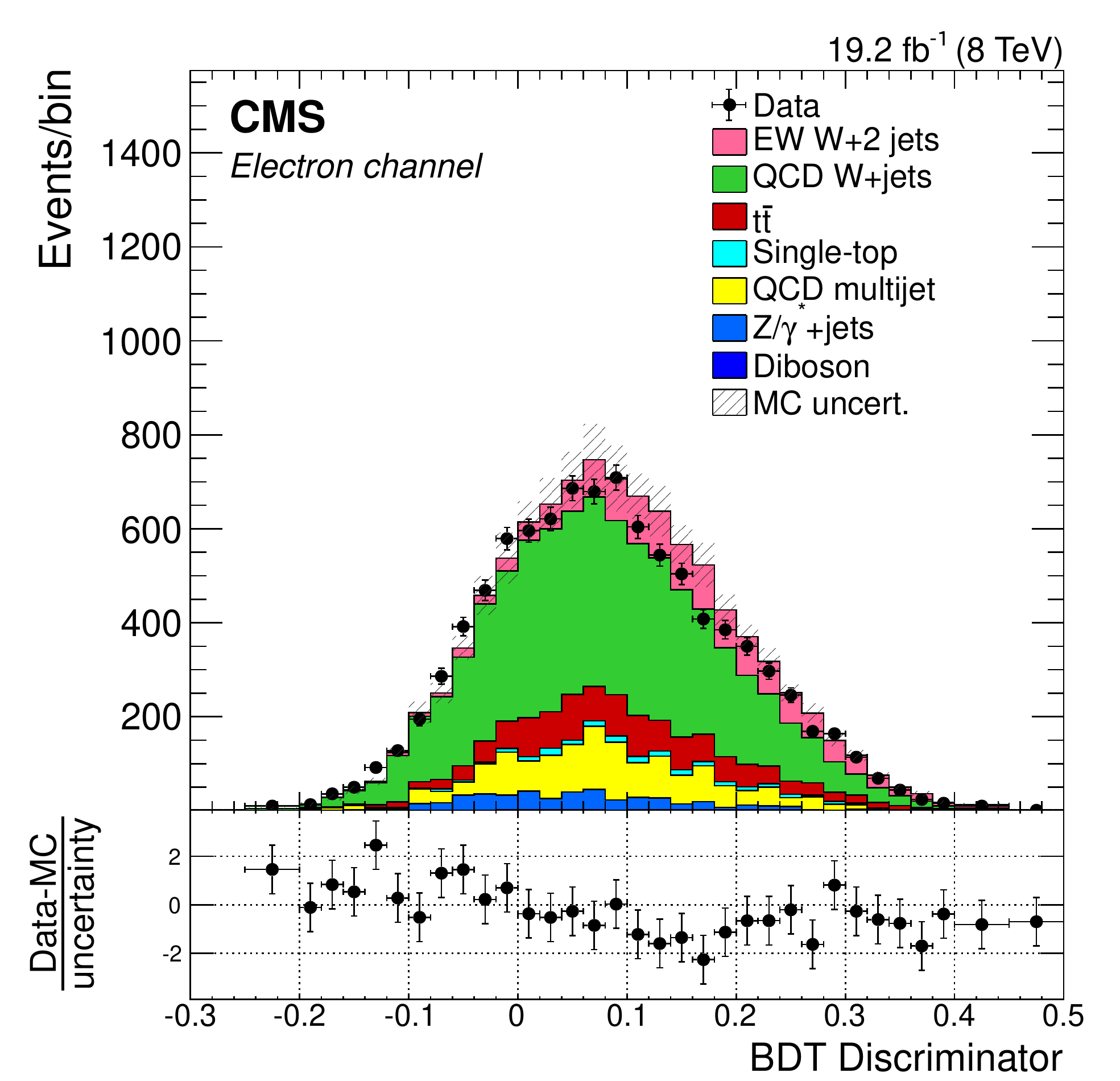}
\includegraphics[width=.49\textwidth]{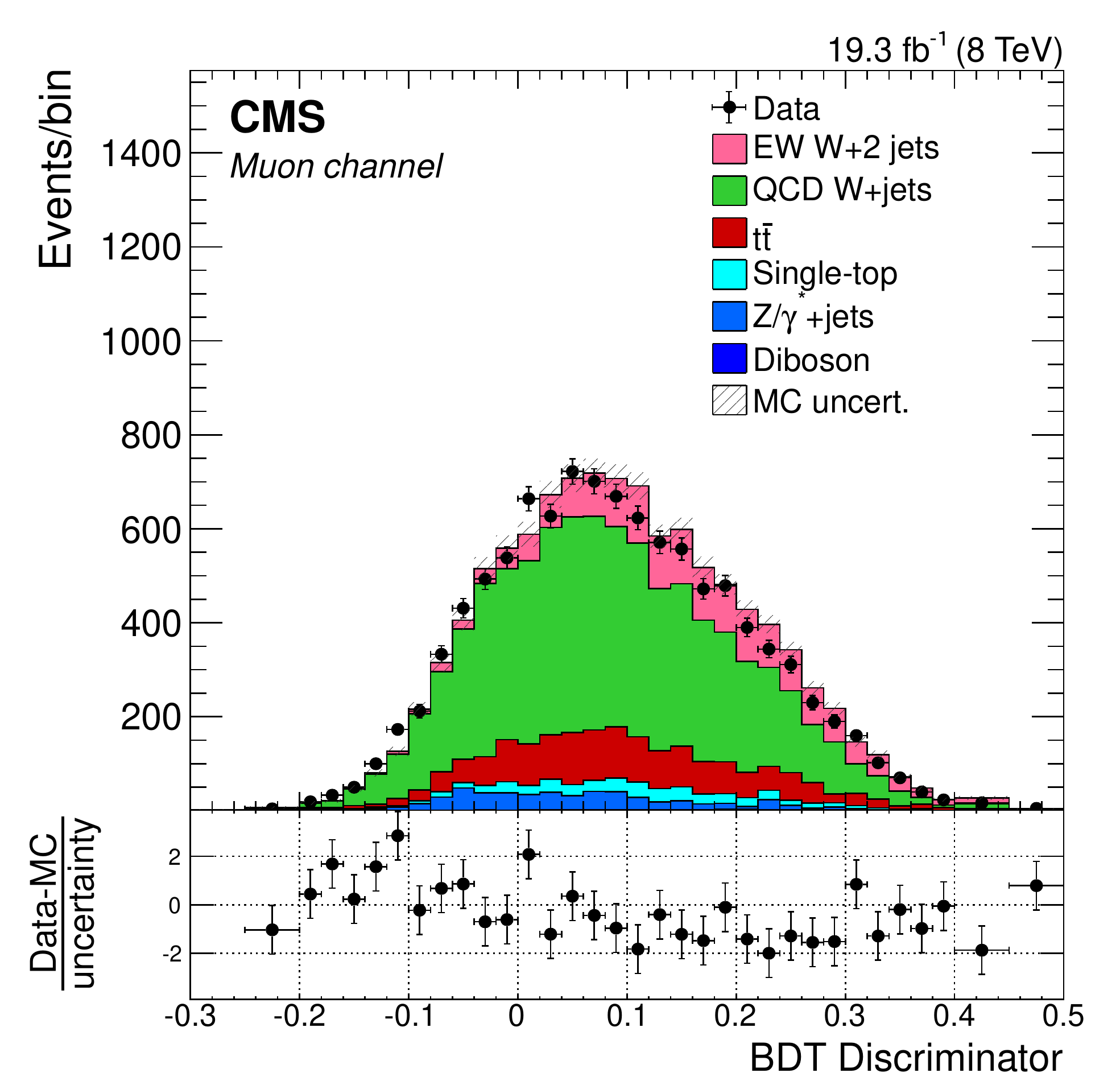}
}
\caption{
Comparison of the distributions of BDT discriminant output for data and simulation in the electron (left) and muon (right) channels with $\mjj > 1000$\GeV. The simulated QCD $\PW$+jets sample is multiplied by the normalization scale factors and the simulated signal sample is multiplied by the signal strength.
}
\label{fig:BDToutputdistribution}
\end{figure}

The BDT is trained with simulated samples to discriminate the EW $\PW(\to\Pe\nu,\, \Pgm\nu,\, \tau\nu)$+2-jets
signal from the QCD $\PW$+jets events, which is the main background;
selected simulated events with $\mjj > 260\GeV$ are used.
The lower required \mjj value provides a larger sample of events that help
avoid over-training the BDT.
Studies give no indication of a bias as a function of the minimum \mjj.
On average the EW events have higher BDT values
than the background events.
The QCD $\PW$+jet simulation overestimates the event yield in the data set, and therefore
a data-to-MC scaling factor is extracted as follows.
The other background contributions are fixed to their simulated yields while the normalization of the QCD W+jets simulated events is scaled so that the
total number of events in simulation equals that in the data set for BDT values less than 0.1.
The fit is performed for events with $\mjj > 1000\GeV$.
The normalization uncertainties of the other backgrounds are considered
as a source of systematic uncertainty.
Systematic uncertainties related to the remaining discrepancy between data and simulation and to the contamination from signal in the BDT ${<}0.1$ region, especially in the muon channel as shown in Fig.~\ref{fig:BDToutputdistribution}, are discussed in Section~\ref{sec:Systematic}.
The resulting $\PW$+jets normalization factors
are
$0.71 \pm 0.02\stat \pm 0.03\syst$
and
$0.70 \pm 0.02\stat \pm 0.05\syst$
for the electron and muon channels, respectively.
These normalization factors are applied in addition to the NNLO K factor of approximately 1.24, mentioned in Section~\ref{sec:SimulatedSamples}.
The scale factor relative
to LO is thus around 0.87, which agrees with the result in~\cite{Campbell:2003hd} and has also been verified with MCFM. Finally,
similar numbers are found in another CMS analysis involving the VBF
topology~\cite{Chatrchyan:2014tja}.
The QCD $\PW$+jets normalization scale factors are used later in the fit to the $\mjj$ distributions when extracting the number of signal events.

Figure \ref{fig:BDToutputdistribution} shows the BDT output distributions for the electron and muon channels, with the simulated QCD $\PW$+jets sample multiplied by the normalization scale factors described above and the simulated signal sample multiplied by the signal strength we obtain
later in this section.
The uncertainty band includes the statistical uncertainty of the simulated samples, the integrated luminosity uncertainty, and
a systematic uncertainty in the QCD background due to misidentified electrons (to be discussed below).
The distributions of
the difference between data and simulation divided by the uncertainty,
where the uncertainty is computed at each point by combining the uncertainties in data and in simulation, are also shown in these plots.

\begin{figure}[tb]
  {\centering
    \includegraphics[width=0.49\textwidth]{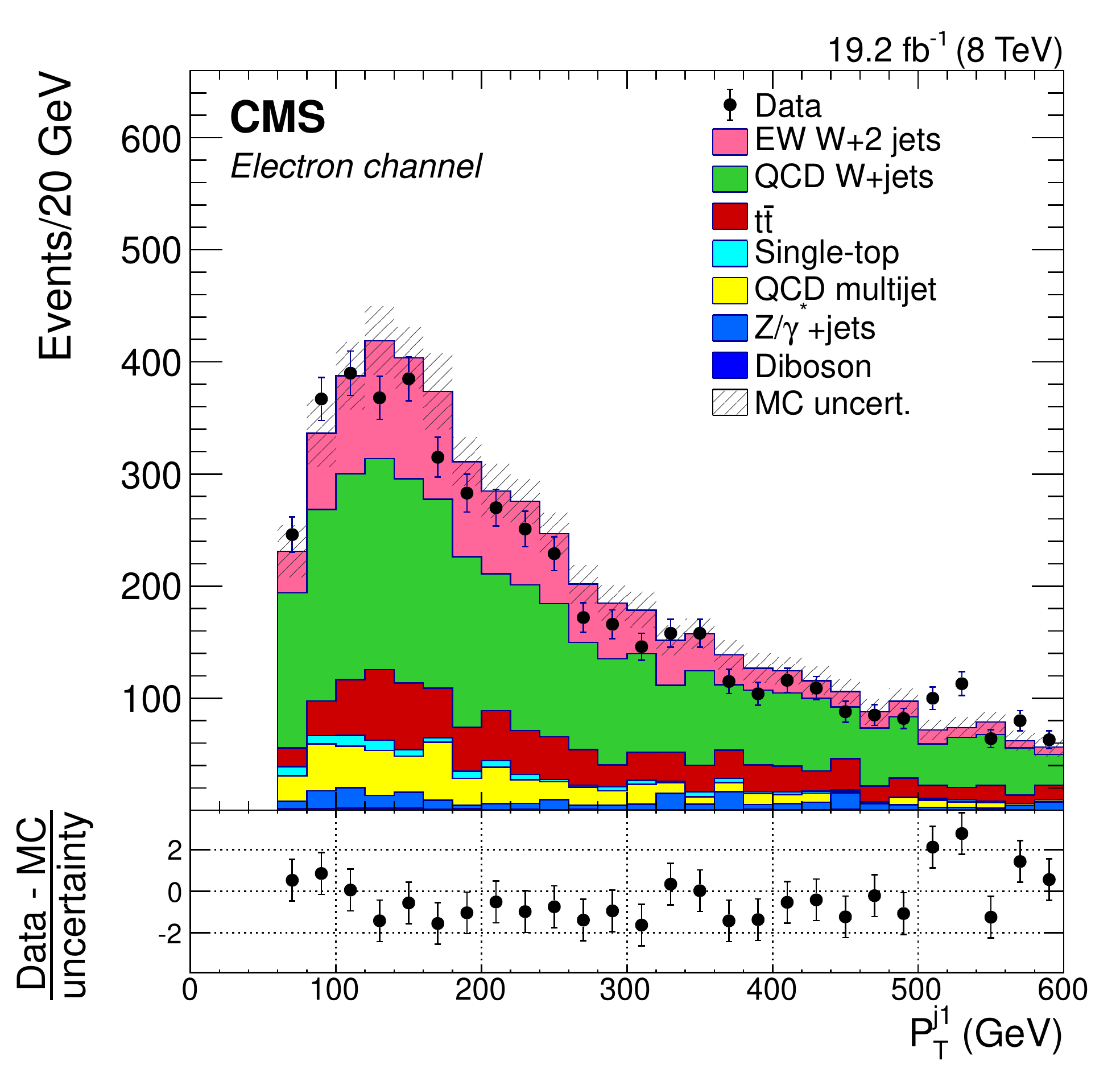}
    \includegraphics[width=0.49\textwidth]{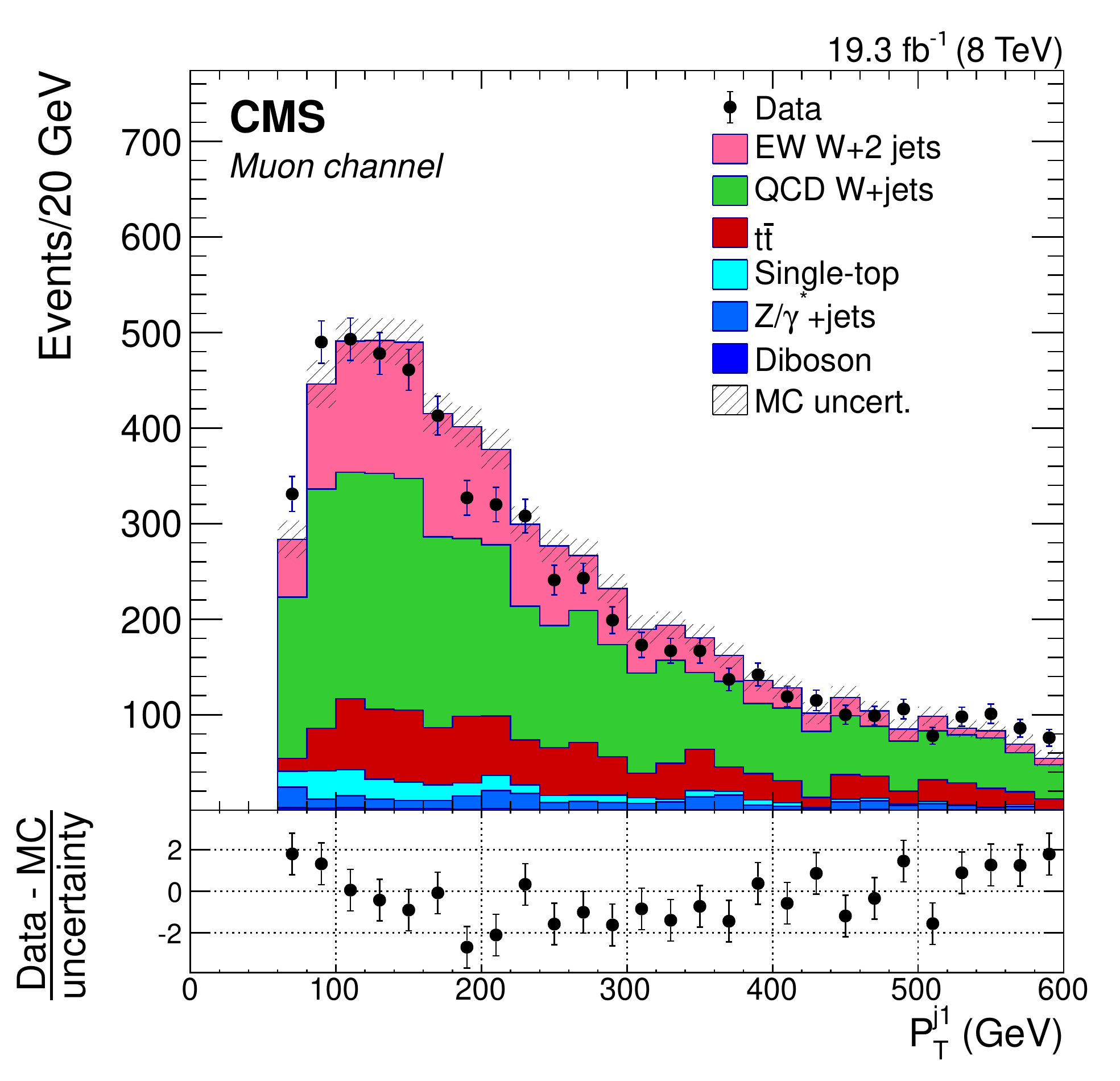}
    \caption{Comparison of the distributions of the leading jet \pt between data and simulation for the electron (left) and muon (right) channels with $\mjj > 1000$\GeV.}
    \label{fig:mu_jet1_pt}}
\end{figure}

\begin{figure}[tb]
  {\centering
    \includegraphics[width=0.49\textwidth]{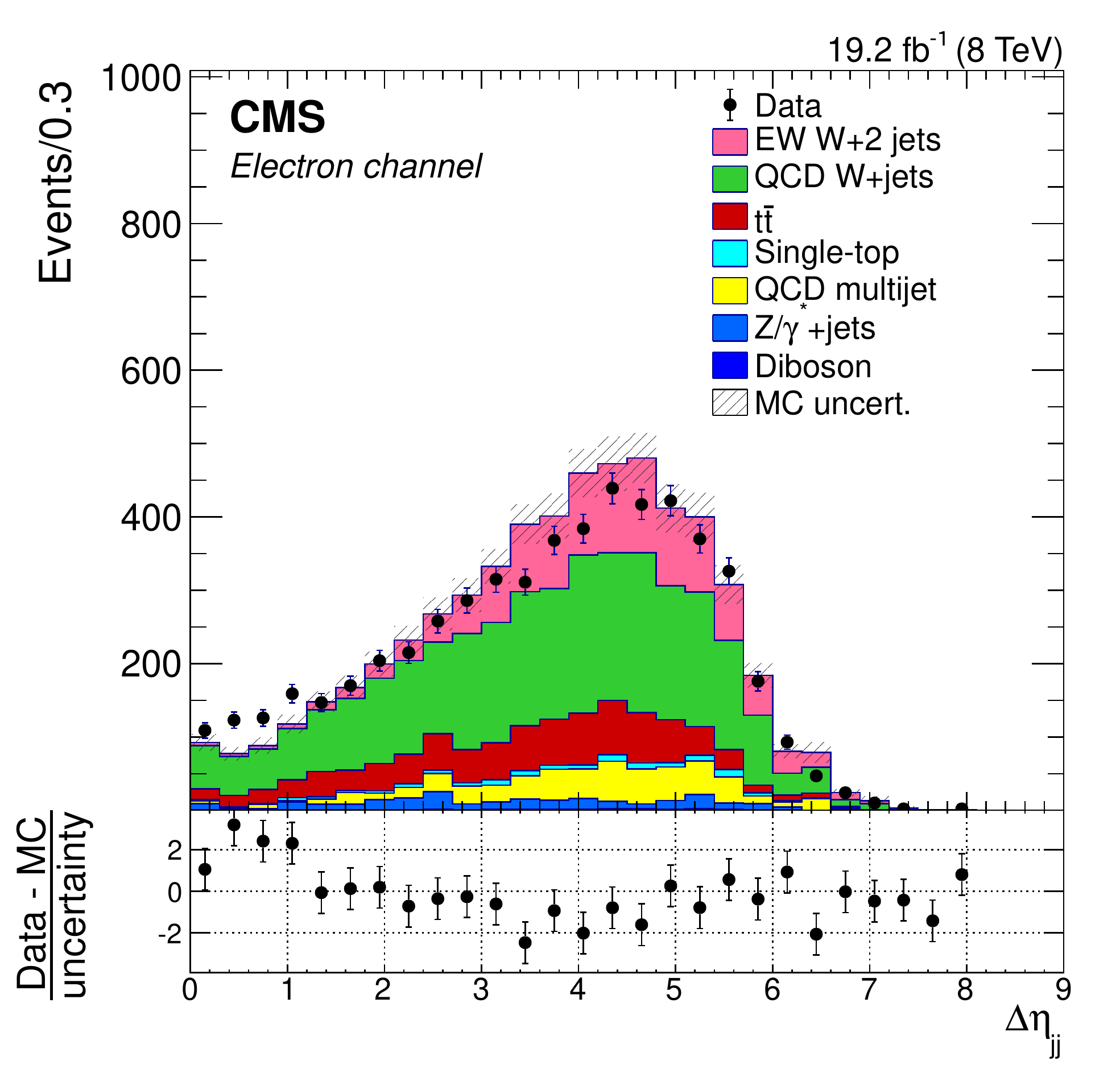}
    \includegraphics[width=0.49\textwidth]{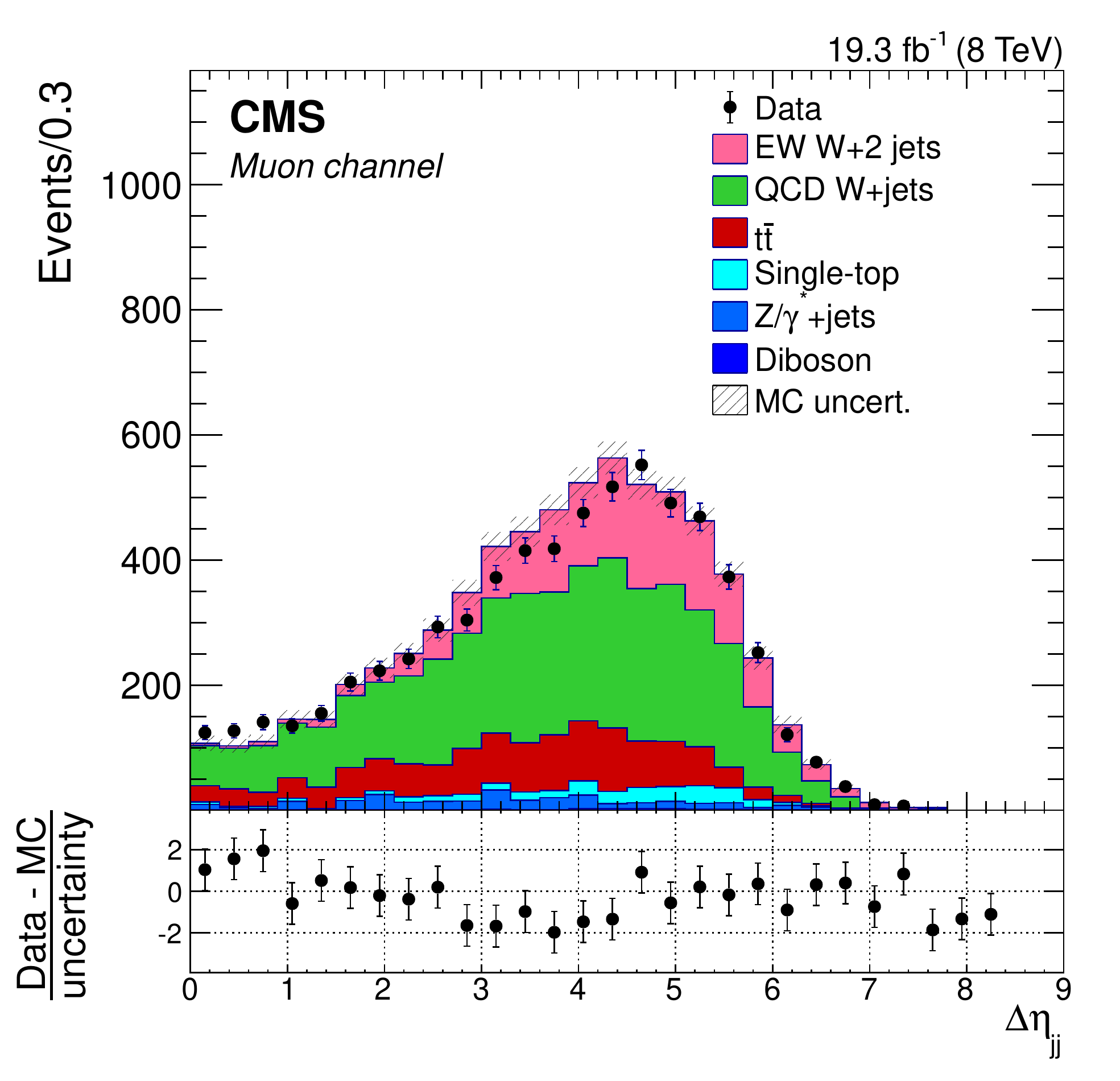}
    \caption{Comparison of the distributions of $\Delta\eta$ between the two leading jets between data and simulation for the electron (left) and muon (right) channels with $\mjj > 1000$\GeV.
      }
    \label{fig:mu_tagjet_deltaeta}}
\end{figure}

The distributions of the leading jet \pt and the $\Delta \eta$ between the two jets for data and simulation after the lepton and jet selection requirements are shown in
Figs.~\ref{fig:mu_jet1_pt} and~\ref{fig:mu_tagjet_deltaeta}, respectively,
with the QCD $\PW$+jets MC sample multiplied by the normalization scale factors described above.

Taking into account the QCD $\PW$+jet background normalization, the extraction of the signal yield is performed using an unbinned maximum-likelihood fit to the $\mjj$ distribution in data.
The signal strength $\mu_{\text{signal}}$, which is defined as the ratio of the extracted signal yield and the expected yield predicted by the LO SM calculation, is the free parameter in the fit.

We employ a parametric function to model the $\mjj$ distributions for the
signal and each of the background contributions listed in Table~\ref{tab:yields}.
Only dijet masses greater than 1000\GeV are used in the fit, as indicated by the study in Ref.~\cite{Rainwater:1996ud}.
Based on the simulation we expect the $\mjj$ distributions to be well-described by the two parameter power law function

\begin{equation}
	\mathcal{F} =  \frac{1}{ {\mjj}{}^{ a_{0} + a_{1} \ln{ (\mjj / 8000 )} } },
\end{equation}

where $a_{0}$ and $a_{1}$ are obtained from the simulation and $\mjj$ is in $\GeV$.
Separate fits are performed to the simulated distribution for the signal and each of the background contributions.
This function provides a good description of the signal and background shapes.

The normalization of the EW $\PW(\to \Pe\nu,\, \Pgm\nu,\, \tau\nu)$+2-jets
contribution
is a free parameter in the fit to the $\mjj$ distribution,
while the shape parameters are fixed to the MC prediction.
The effect of QCD NLO corrections for the EW $\PW(\to \ell\nu)$+2-jets process is tested by modifying the shape parameters in order to reproduce the prediction of \textsc{vbfnlo}~2.6~\cite{VBFNLO}; the resulting variation in the evaluated signal strength is below 1\%.

The QCD $\PW$+jets background shape parameters are left free during the $\mjj$ fit to data because of
the poor agreement between data and simulation for this background. The normalization of the QCD $\PW$+jets background is fixed to the fit result from the BDT distribution, as described above.

Multijet events can be misidentified as signal because of the nonnegligible probability of jets to be misidentified as electrons.
The fraction of fake electrons in the single electron data passing the selection described in Section~\ref{sec:EventSelection} is obtained from an independent two-component fit to the $\MET$ distribution following Ref.~\cite{VBTF}.
The normalization and shape parameters are fixed in the fit.
We conservatively assume a 50\% uncertainty in the QCD multijet yield when fitting the data.

The top quark background is a combination of \ttbar\ and single top quark processes with the simulated samples
normalized according to
the known cross sections. The shape parameters are obtained from the simulation and fixed during the fit.
The top quark background normalization is assumed to have a Gaussian probability density function
with a width of
7\%~\cite{Czakon:2011xx}.

Other background processes, such as diboson production and DY, are minor and are represented by their components with the corresponding normalization fixed in the fit.
The shape parameters are obtained from simulation and are also fixed.

\begin{figure*}[t]
\includegraphics[width=0.32\columnwidth]{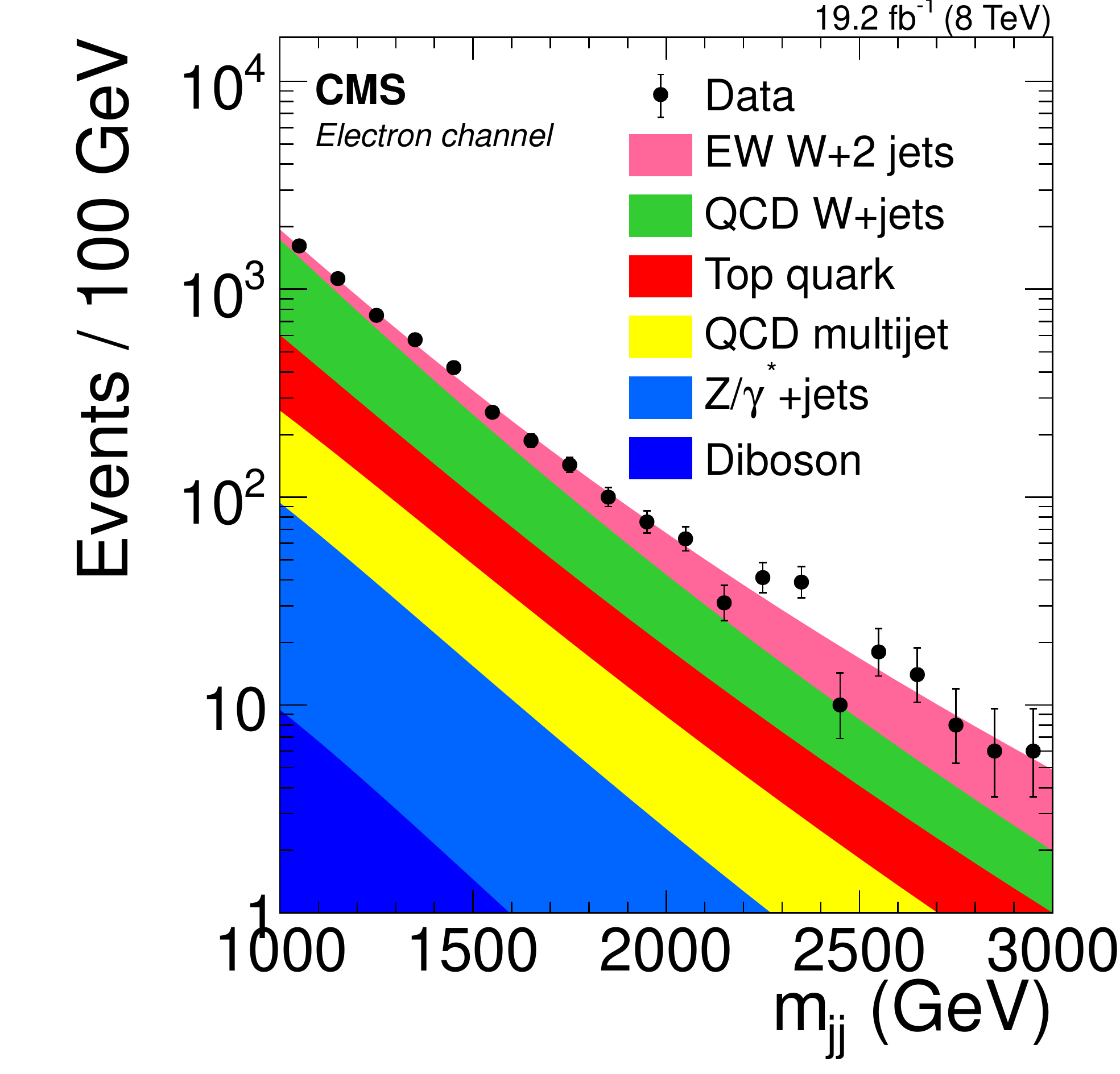}
\includegraphics[width=0.32\columnwidth]{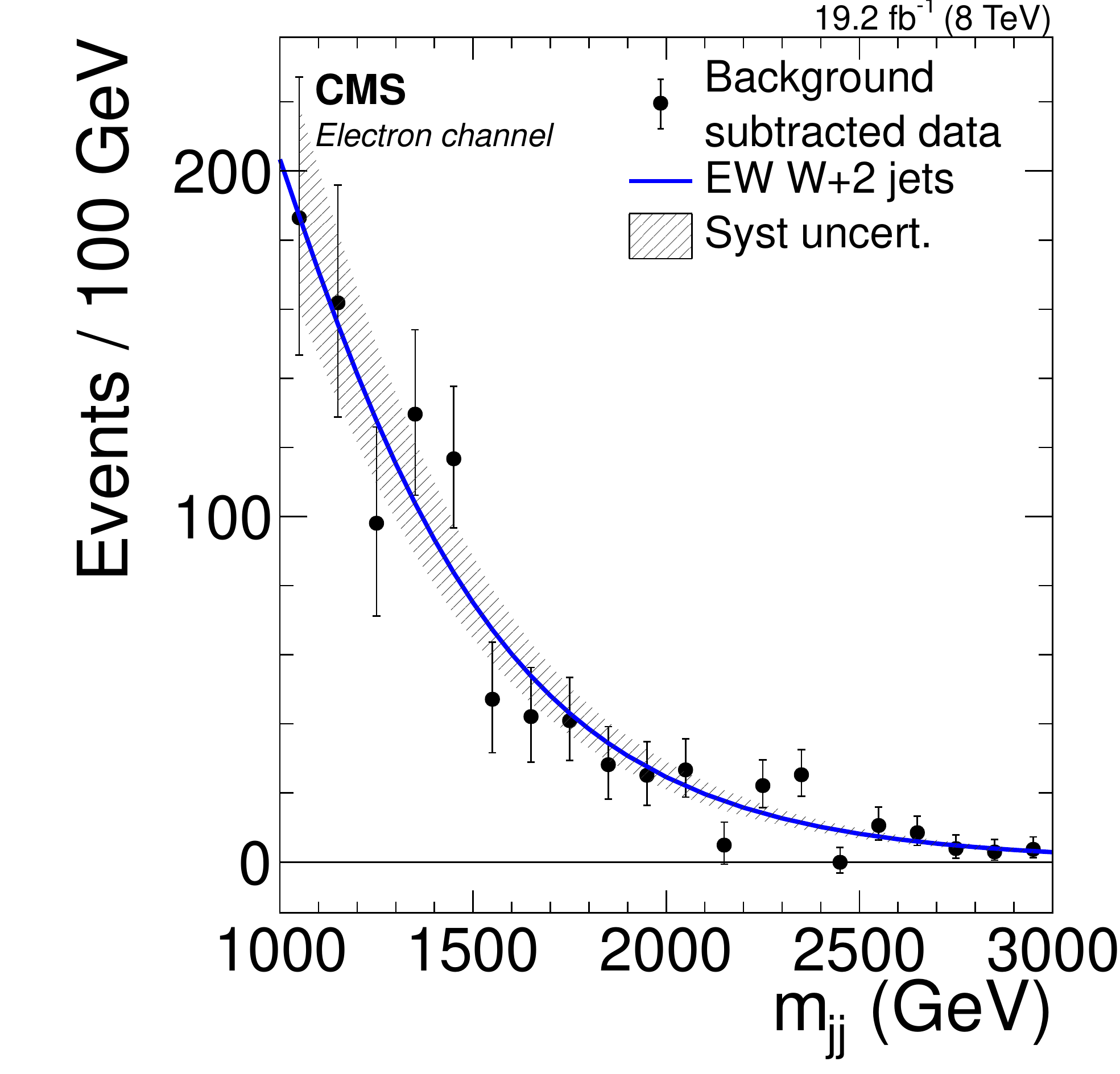}
\includegraphics[width=0.32\columnwidth]{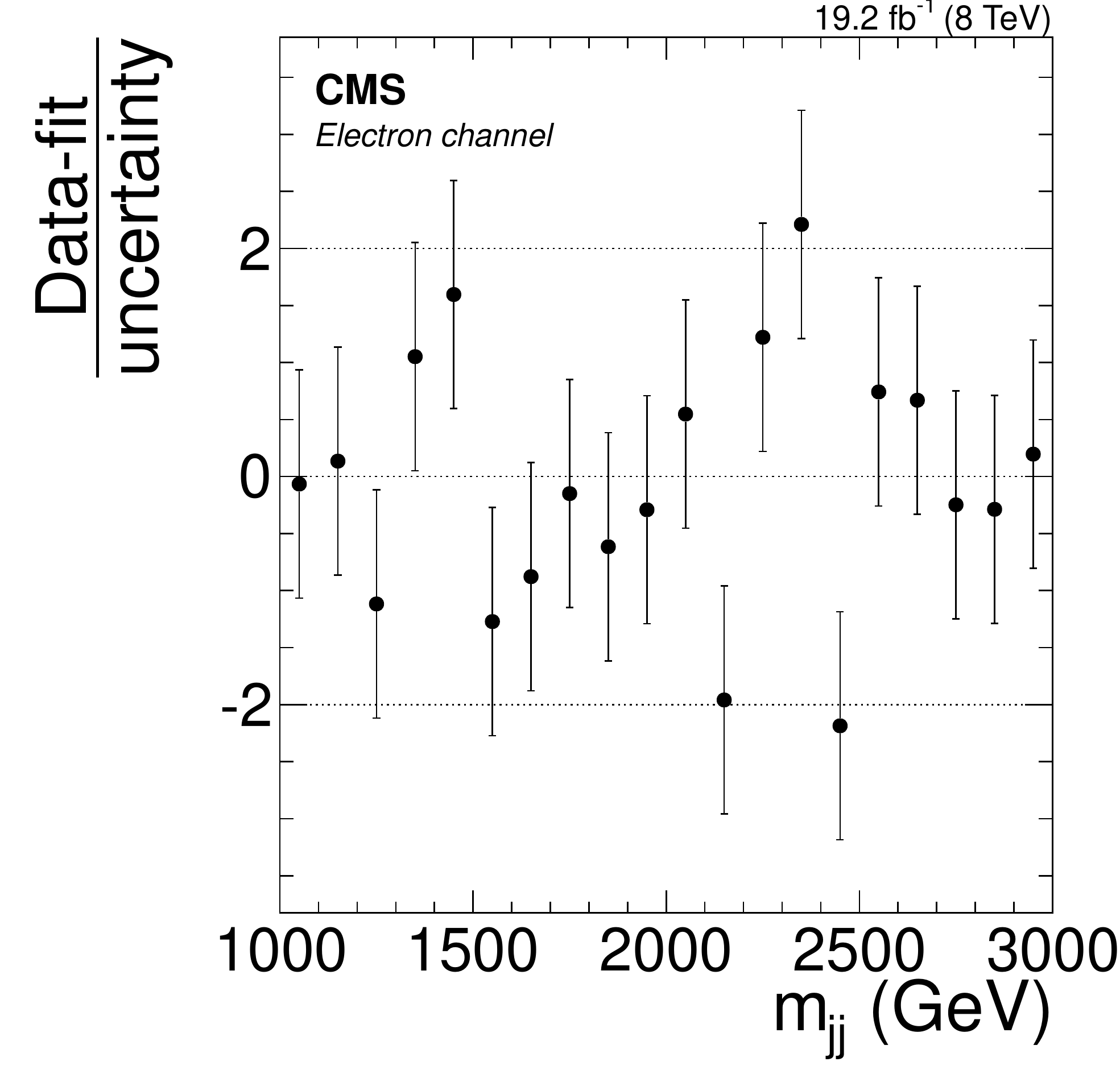}\\
\includegraphics[width=0.32\columnwidth]{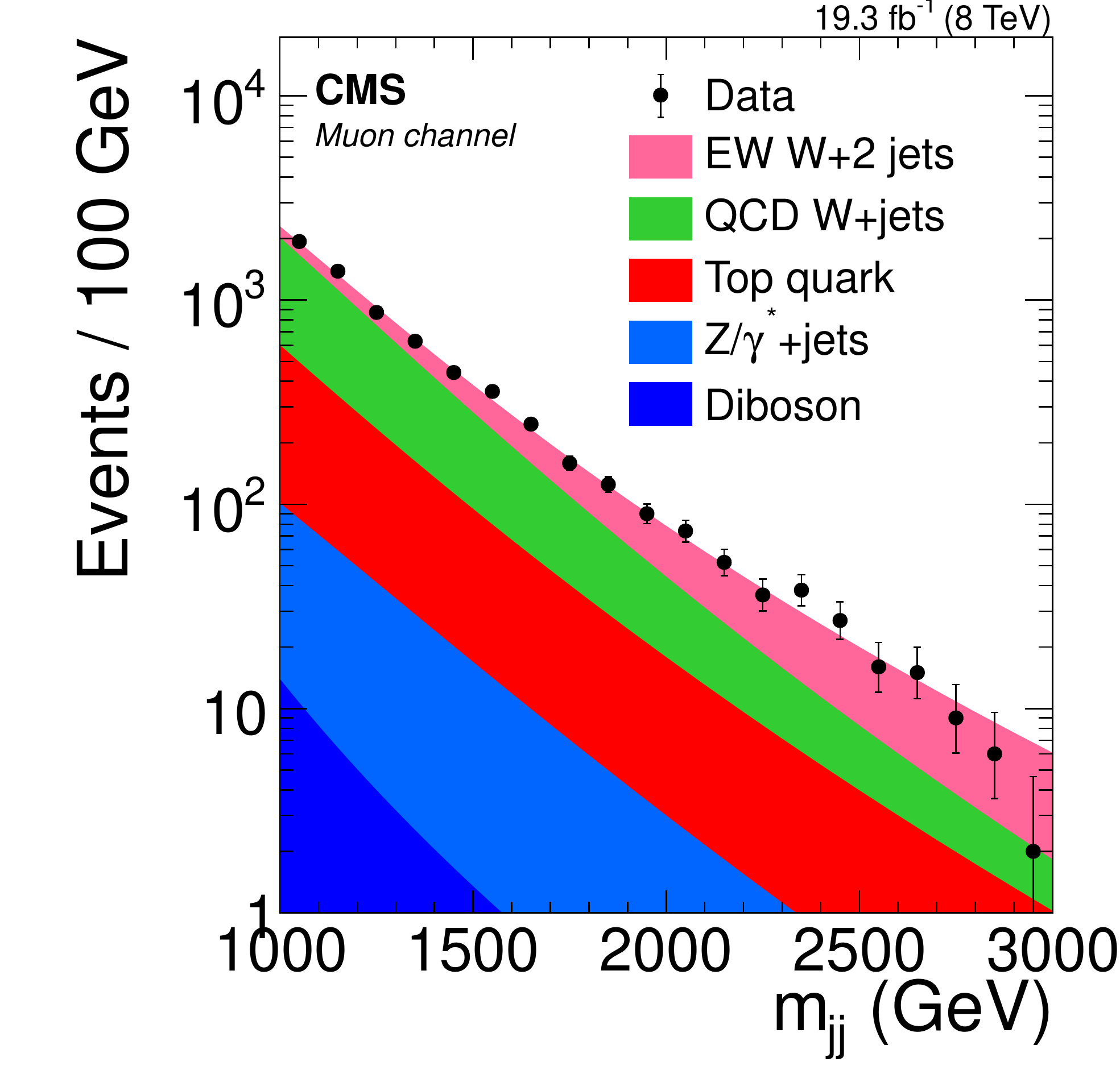}
\includegraphics[width=0.32\columnwidth]{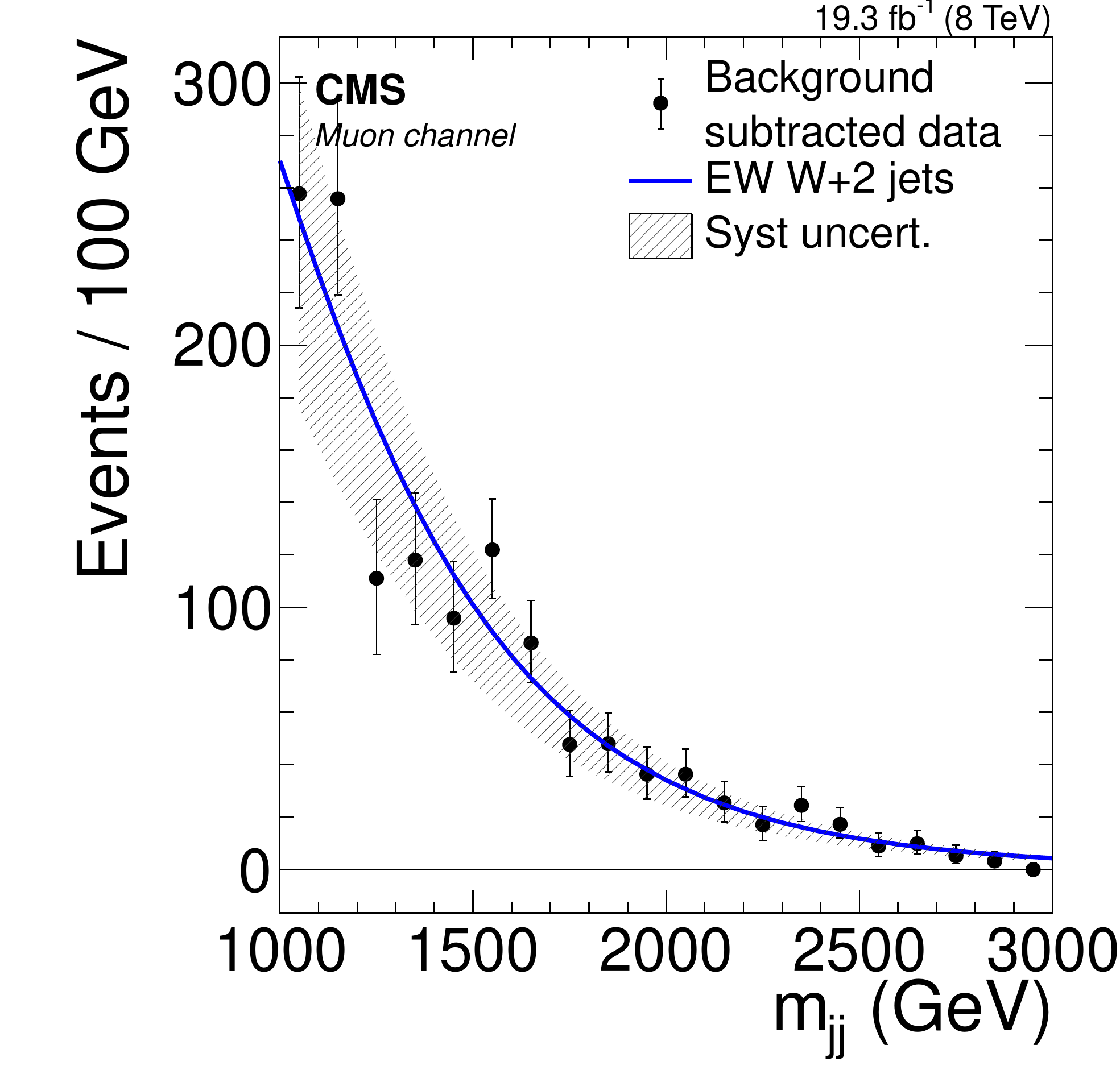}
\includegraphics[width=0.32\columnwidth]{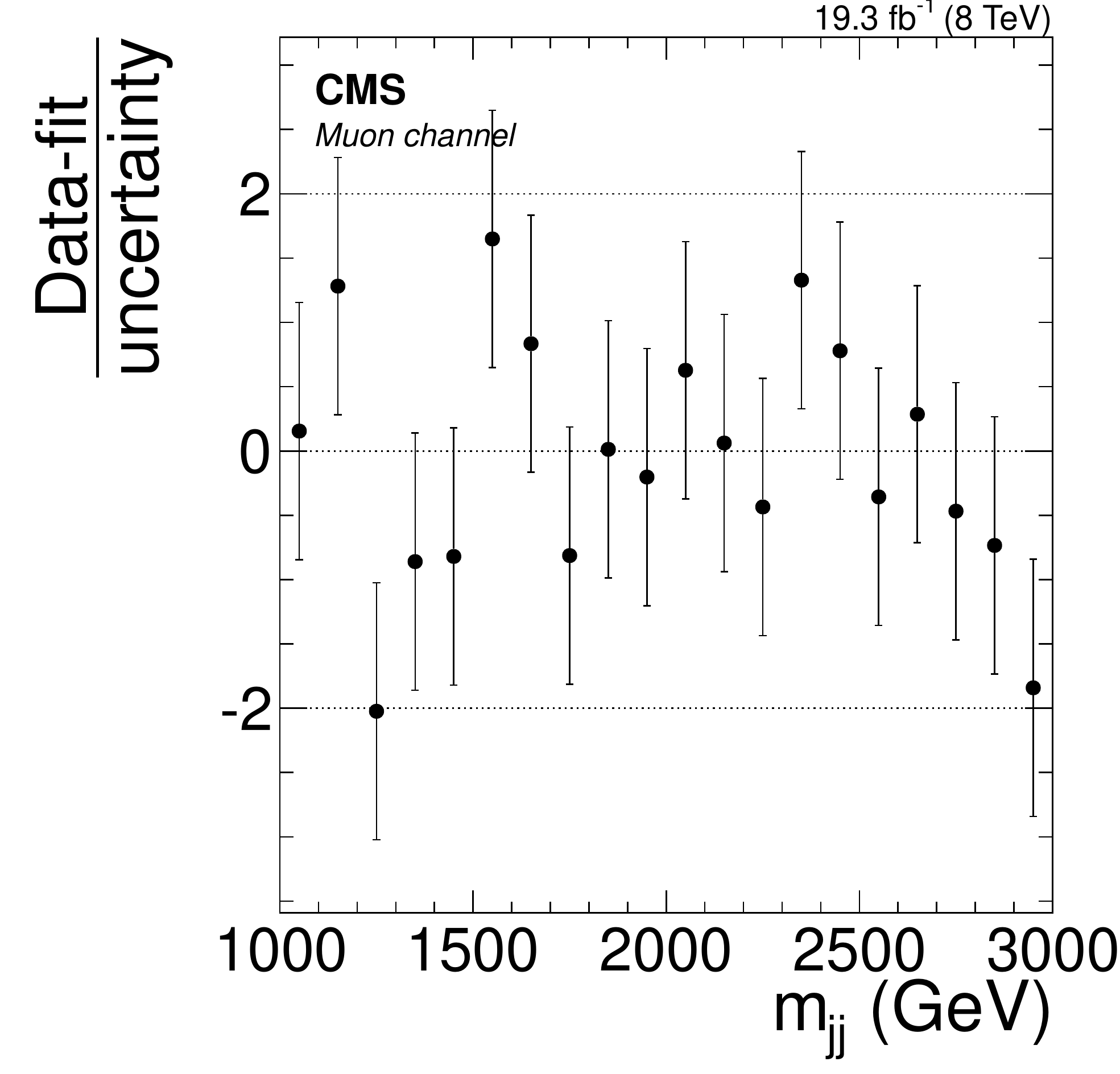}
\caption{
Distribution of the two-jet invariant mass \mjj for electron (upper) and muon (lower) channels.
Fitted projections of signal and background processes are plotted as shaded regions (left plots).
The \mjj distributions are shown after subtraction of all components except the EW $\PW$+2-jets process (center plots).
Finally, the $(\text{data} - \text{fit})/\text{uncertainty}$ distributions are shown (right plots).
Here the error bars represent the statistical uncertainties of the data.
}
\label{fig:Fig1}
\end{figure*}
\begin{table*}[htbp]
	\begin{center}
  \topcaption{The expected event yields and the ratio of the measured and expected yields extracted from the maximum-likelihood fit to data.}
  \label{tab:yields}
\begin{tabular}{l l r l r }
\hline
Process            &  \multicolumn{2}{c}{Electrons} & \multicolumn{2}{c}{Muons} \\
 \hline
 Data & \multicolumn{2}{c}{5481} & \multicolumn{2}{c}{6514} \\
\hline
& Pre-fit &  \begin{tabular}[l]{@{}l@{}}Measured\\ratio\end{tabular}  &  Pre-fit &  \begin{tabular}[l]{@{}l@{}}Measured\\ratio\end{tabular} \\
               \hline
$\PW$+jets         & 3913   &  0.71 (fixed)               & 5084   &  0.70 (fixed)       \\
Top quark      & 933    &   1.00 $\pm$ 0.07          & 1357   & 1.00 $\pm$ 0.07    \\
QCD multijet       &  510   &  \begin{tabular}[l]{@{}l@{}}fixed to\\\MET fit\\in data\end{tabular}  &  ---   &   ---              \\
DY+jets         & 236    &  1.00 (fixed)               & 256    &  1.00 (fixed)       \\
Diboson       & 26     &  1.00 (fixed)               & 29     & 1.00 (fixed)        \\
 \hline
 Total backgrounds & \multicolumn{2}{c}{4488} & \multicolumn{2}{c}{5179} \\
 \hline
 Data $-$ backgrounds & \multicolumn{2}{c}{993} & \multicolumn{2}{c}{1335}\\
 \hline
EW $\PW$+2-jets    & 1195   &  0.83 $\pm$ 0.08           & 1541   & 0.87 $\pm$ 0.08    \\
 \hline
\end{tabular}
\end{center}
\end{table*}

Figure \ref{fig:Fig1} (left) shows the observed $\mjj$ distributions for
the electron and muon channels, together with the fitted projections of the
contribution of the signal and background processes
described by the two-parameter power law function.
The shapes of the different fit components look similar,
nonetheless, their slopes are different.
Figure \ref{fig:Fig1} (center)
shows the \mjj distribution after subtracting all SM background contributions.
Figure \ref{fig:Fig1} (right) presents the distributions
of the difference between the \mjj values in data and from the fit, divided by the data uncertainty.
The yields of the various SM components, as determined by the fit,
are reported in Table~\ref{tab:yields}.
The resulting signal strengths $\mu_{\text{signal}}$ are $0.83 \pm 0.08\stat$ and $0.87 \pm 0.08\stat$ for electron and muon channels, respectively.
While a simple counting of events in Table~\ref{tab:yields} can determine the number of EW $\PW$+2-jets events, the fitting procedure includes the distribution of Fig.~\ref{fig:Fig1} and allows us to use the shape parameters when calculating event yields.

This approach produces
an acceptable
model of the data (Fig.~\ref{fig:Fig1} left)
and allows us to extract the EW $\PW$+2-jets signal (Fig.~\ref{fig:Fig1} center).
In Fig.~\ref{fig:Fig1} right the mean value of the pull distribution is $-0.04$ (0.03), consistent with zero in the electron (muon) channel, and the pull variance is 0.90\,(0.87), consistent with unity in the electron (muon) channel.

\section{Systematic uncertainties}
\label{sec:Systematic}

Systematic uncertainties arising from the jet energy scale and resolution are estimated by varying the calibration parameters up and down by one standard deviation in the simulation and evaluating the impact on the cross sections~\cite{Chatrchyan:2011ds}.

A different parametric function is used to model the QCD $\PW$+jets shape to estimate the potential influence of the choice of parametric function in the fit result, and it is considered as a source of systematic uncertainty.
To estimate the systematic uncertainty from the disagreement between data and simulation in the BDT output distributions, we divide the sideband in BDT ($\mathrm{BDT} < 0.1$) used to estimate the scale factor for the QCD $\PW$+jets normalization into two roughly equally populated regions ($\mathrm{BDT}< 0.02$, $0.02 < \mathrm{BDT} < 0.1$).
We recompute the scale factor using these two subsamples and
propagate the variation in the normalization to the final result.
The uncertainty with respect to the QCD $\PW$+jets normalization is 4.9\% (7.1\%) in the electron (muon) channel.
The uncertainty propagated to the cross section is 12.9\% (16.6\%) in the electron (muon) channel.
The uncertainty on the normalization is then added in quadrature with the shape uncertainty as shown in Table~\ref{tab:signalSyst}.

Background $\ttbar$ samples produced with different generators and parton shower models
(\MCATNLO + \HERWIG and \POWHEG + \PYTHIA)
are used as alternative to the default $\ttbar$ \MADGRAPH shapes.
We also select a $\ttbar$ enriched phase space to compare the $\ttbar$ sample produced with \MADGRAPH with data.
Reasonable agreement
is found between the \MADGRAPH \ttbar sample and data.

The EW $\PW(\to \ell\nu)$+2-jets
and QCD $\PW(\to \ell\nu)$+2-jets
processes have a positive interference term, which is neglected.
In order to estimate the effect of the interference, two additional samples are generated using \MADGRAPH:
\begin{itemize}
	\item
		QCD $\PW(\to\ell\nu)$+2-jets
		sample (strong process only),
	\item
		EW $\PW(\to \ell\nu)$+2-jets
		+ QCD $\PW(\to \ell\nu)$+2-jets
		+ interference effect sample (mixture of EW and strong processes).
\end{itemize}
We subtract the QCD and EW processes from the mixture sample to estimate the effect of the interference.
The contribution of the interference effect is considered as an additional background with a fixed shape and normalization.
The true EW signal strength is ${\approx}12\%$ smaller than the apparent signal strength because the interference is positive.

The fraction of jets faking electrons
is varied by ${\pm}50\%$ in the electron channel to estimate the uncertainty.
A small difference in \MET resolution~\cite{Khachatryan:2014gga} between data and simulation affects the signal acceptance at the 0.5\% level.
We also consider systematic uncertainties in the trigger efficiency (1\%) and in the lepton reconstruction and
selection efficiencies (2\%)~\cite{VBTF}.
The uncertainty in the luminosity measurement is 2.6\%~\cite{lumiPAS}.
Finally, we calculated the acceptance for the fiducial region with {\HERWIGpp} signal samples, resulting in a 1.7\% change in the measured cross section.

The systematic uncertainties are summarized in Table~\ref{tab:signalSyst}.

 \begin{table}[thb!]
   \begin{center}
	\topcaption{\label{tab:signalSyst}Sources of systematic uncertainties and the magnitude of their effect on the fiducial cross section.}
   \begin{tabular}{l|c|c}
  \hline
  Source of uncertainty                        & Electrons & Muons   \\
  \hline
  Integrated luminosity                        & \multicolumn{2}{c}{2.6\%}   \\
  \hline
  Jet energy scale                             & \x5.4\%     & \x7.3\%   \\
  Jet energy resolution                        & \x2.2\%     & \x3.7\%   \\
  QCD $\PW$+jets shape and normalization               & 13.0\%    & 16.7\%  \\
  Top quark background shape and normalization            & \x5.5\%     & \x6.0\%   \\
  Interference effect                          & 14.4\%    & 13.8\%  \\
  Jets faking electrons fraction (electron channel)   & \x4.4\%     & ---    \\
  Lepton trigger efficiency                    & \x0.9\%     & \x1.0\%   \\
  Lepton selection efficiency                  & \x1.8\%     & \x2.0\%   \\
  Pileup                                       & ${<}1\%$    & ${<}1\%$  \\
  Fiducial acceptance                          & \x1.7\%     &  \x1.7\%  \\
  \hline
  Total (without integrated luminosity)                   & 21.6\%    & 24.1\%  \\
  \hline
   \end{tabular}
   \end{center}
 \end{table}

\section{Results}

The fiducial EW $\PW$+2-jets cross section is calculated for $\PW$~bosons decaying to electrons or muons and for $\ptjone > 60$\GeV, $\ptjtwo > 50$\GeV, $\abs{\etaj} <  4.7$, and $\mjj >  1000$\GeV. Here we include neutrinos in our definition of generator jets, which are reconstructed in the same manner as the PF jets mentioned above, but now at generator level.

The fiducial cross section is computed as
\begin{equation}
	\sigma_{\text{fiducial}} =\sigma_{\text{generator}} \mu_{\text{signal}} \epsilon_{\text{acceptance}} ,
\end{equation}
where $\sigma_{\text{generator}}$ is the generator level cross section,
$\mu_{\text{signal}}$ is the signal strength,
and $\epsilon_{\text{acceptance}}$ is the acceptance evaluated with the EW $\PW(\to\Pe\nu, \Pgm\nu, \tau\nu$)+2-jets sample.

To calculate $\epsilon_{\text{acceptance}}$, we divide the number of events in the fiducial region by the total number of generated events
which results in an value of 0.0448.
The uncertainty due to the limited size of the simulated sample is 0.7\%.
The fiducial selections are different from the selections used for the measurement
(as summarized in Table~\ref{tab:SELECTION}),
for which the acceptance is 0.0128.
The uncertainty due to the limited size of the simulated sample is 1.3\%.

We use the best linear unbiased estimate
method~\cite{Lyons:1988rp,Valassi:2003mu}
to combine the results in the electron and muon channels, assuming that the statistical uncertainties are uncorrelated and the systematic
uncertainties are 100\% correlated between the two channels.
Because in this case the (fully correlated) systematic uncertainties are much larger than the statistical uncertainties, the method yields
a statistical uncertainty for the combined result that is not much smaller than the statistical uncertainties for the individual channels.
The results are summarized in Table~\ref{tab:measuredCrossSection},
and are in agreement
with the SM LO prediction of
$0.50 \pm 0.02\,(\text{scale}) \pm 0.02\,(\mathrm{PDF})\unit{pb}$
obtained from {\MADGRAPH}5\_a\MCATNLO~2.1~interfaced to \PYTHIA 6.4.
The statistical significance of the observation is approximately four standard deviations.

\begin{table}[h]
\begin{center}
	\topcaption{\label{tab:measuredCrossSection}
		The measured values for the EW $\PW(\to \ell\nu)$+2-jets
		fiducial cross section.
	}
  \begin{tabular}{l c}
    \hline
    Channel &  Measured cross section\\
    \hline
	 Electron           &  $0.41 \pm 0.04\stat \pm 0.09\syst \pm 0.01\lum\unit{pb}$\\
    Muon        &  $0.43 \pm 0.04\stat \pm 0.10\syst \pm 0.01\lum\unit{pb}$\\
	Combined  & $0.42 \pm 0.04\stat \pm 0.09\syst \pm 0.01\lum\unit{pb}$\\
    \hline
  \end{tabular}
\end{center}
\end{table}

\section{Summary}
A measurement has been performed of the cross section for the electroweak production of $\PW$~bosons produced in association with two forward jets in proton-proton collisions.
The $\PW$ bosons were identified
through their decay to electrons and muons.
The data set was collected by the CMS
experiment and corresponds to
an integrated luminosity of 19.2\,$(19.3) \pm0.5$\fbinv
in the electron (muon) channel at $\sqrt{s}=8\TeV$.
The measured value of the fiducial electroweak $\PW$+2-jets cross section, for $\PW$~bosons decaying to electrons or muons and for $\ptjone > $ 60\GeV, $\ptjtwo > $ 50\GeV, $\abs{\etaj} < $ 4.7, and $\mjj > $ 1000\GeV, is $0.42 \pm 0.04\stat \pm 0.09\syst \pm 0.01\lum \unit{pb}$,
consistent with the SM LO prediction of
$ 0.50 \pm 0.02\,(\text{scale}) \pm 0.02\,(\mathrm{PDF})\unit{pb}$
obtained via {\MADGRAPH}5\_a\MCATNLO~2.1~interfaced with \PYTHIA 6.4.
This is the first
measurement of the cross section for electroweak $\PW$+2-jets production.

\clearpage

\begin{acknowledgments}

We congratulate our colleagues in the CERN accelerator departments for the excellent performance of the LHC and thank the technical and administrative staffs at CERN and at other CMS institutes for their contributions to the success of the CMS effort. In addition, we gratefully acknowledge the computing centers and personnel of the Worldwide LHC Computing Grid for delivering so effectively the computing infrastructure essential to our analyses. Finally, we acknowledge the enduring support for the construction and operation of the LHC and the CMS detector provided by the following funding agencies: BMWFW and FWF (Austria); FNRS and FWO (Belgium); CNPq, CAPES, FAPERJ, and FAPESP (Brazil); MES (Bulgaria); CERN; CAS, MoST, and NSFC (China); COLCIENCIAS (Colombia); MSES and CSF (Croatia); RPF (Cyprus); SENESCYT (Ecuador); MoER, ERC IUT and ERDF (Estonia); Academy of Finland, MEC, and HIP (Finland); CEA and CNRS/IN2P3 (France); BMBF, DFG, and HGF (Germany); GSRT (Greece); OTKA and NIH (Hungary); DAE and DST (India); IPM (Iran); SFI (Ireland); INFN (Italy); MSIP and NRF (Republic of Korea); LAS (Lithuania); MOE and UM (Malaysia); BUAP, CINVESTAV, CONACYT, LNS, SEP, and UASLP-FAI (Mexico); MBIE (New Zealand); PAEC (Pakistan); MSHE and NSC (Poland); FCT (Portugal); JINR (Dubna); MON, RosAtom, RAS and RFBR (Russia); MESTD (Serbia); SEIDI and CPAN (Spain); Swiss Funding Agencies (Switzerland); MST (Taipei); ThEPCenter, IPST, STAR and NSTDA (Thailand); TUBITAK and TAEK (Turkey); NASU and SFFR (Ukraine); STFC (United Kingdom); DOE and NSF (USA).

{\tolerance=1200

Individuals have received support from the Marie-Curie program and the European Research Council and EPLANET (European Union); the Leventis Foundation; the A. P. Sloan Foundation; the Alexander von Humboldt Foundation; the Belgian Federal Science Policy Office; the Fonds pour la Formation \`a la Recherche dans l'Industrie et dans l'Agriculture (FRIA-Belgium); the Agentschap voor Innovatie door Wetenschap en Technologie (IWT-Belgium); the Ministry of Education, Youth and Sports (MEYS) of the Czech Republic; the Council of Science and Industrial Research, India; the HOMING PLUS program of the Foundation for Polish Science, cofinanced from European Union, Regional Development Fund, the Mobility Plus program of the Ministry of Science and Higher Education, the National Science Center (Poland), contracts Harmonia 2014/14/M/ST2/00428, Opus 2013/11/B/ST2/04202, 2014/13/B/ST2/02543 and 2014/15/B/ST2/03998, Sonata-bis 2012/07/E/ST2/01406; the Thalis and Aristeia programs cofinanced by EU-ESF and the Greek NSRF; the National Priorities Research Program by Qatar National Research Fund; the Programa Clar\'in-COFUND del Principado de Asturias; the Rachadapisek Sompot Fund for Postdoctoral Fellowship, Chulalongkorn University and the Chulalongkorn Academic into Its 2nd Century Project Advancement Project (Thailand); and the Welch Foundation, contract C-1845.

}
\end{acknowledgments}

\clearpage

\bibliography{auto_generated}

\cleardoublepage \appendix\section{The CMS Collaboration \label{app:collab}}\begin{sloppypar}\hyphenpenalty=5000\widowpenalty=500\clubpenalty=5000\input{SMP-13-012-authorlist.tex}\end{sloppypar}
\end{document}

%% file: SMP-13-012-authorlist.tex
\textbf{Yerevan Physics Institute,  Yerevan,  Armenia}\\*[0pt]
V.~Khachatryan, A.M.~Sirunyan, A.~Tumasyan
\vskip\cmsinstskip
\textbf{Institut f\"{u}r Hochenergiephysik der OeAW,  Wien,  Austria}\\*[0pt]
W.~Adam, E.~Asilar, T.~Bergauer, J.~Brandstetter, E.~Brondolin, M.~Dragicevic, J.~Er\"{o}, M.~Flechl, M.~Friedl, R.~Fr\"{u}hwirth\cmsAuthorMark{1}, V.M.~Ghete, C.~Hartl, N.~H\"{o}rmann, J.~Hrubec, M.~Jeitler\cmsAuthorMark{1}, V.~Kn\"{u}nz, A.~K\"{o}nig, M.~Krammer\cmsAuthorMark{1}, I.~Kr\"{a}tschmer, D.~Liko, T.~Matsushita, I.~Mikulec, D.~Rabady\cmsAuthorMark{2}, N.~Rad, B.~Rahbaran, H.~Rohringer, J.~Schieck\cmsAuthorMark{1}, R.~Sch\"{o}fbeck, J.~Strauss, W.~Treberer-Treberspurg, W.~Waltenberger, C.-E.~Wulz\cmsAuthorMark{1}
% ...
\vskip\cmsinstskip
% ...
1:~~Also at Vienna University of Technology, Vienna, Austria\\
2:~~Also at CERN, European Organization for Nuclear Research, Geneva, Switzerland\\
% ...
